\documentstyle[twoside,fleqn,espcrc2,psfig]{article}

\title{Mean link versus average plaquette tadpoles
       in lattice NRQCD}
\author{Norman H. Shakespeare and Howard D. Trottier
\address{Physics Department, 
Simon Fraser University, Burnaby, B.C., CANADA V5A 1S6}}

\begin{document}

\begin{abstract}
We compare mean-link and average plaquette tadpole renormalization 
schemes in the context of the quarkonium hyperfine splittings in lattice 
NRQCD. Simulations are done for the three quarkonium systems $c\bar c$, 
$b\bar c$, and $b\bar b$.
The hyperfine splittings are computed both at leading and at 
next-to-leading order in the relativistic expansion. Results are 
obtained at a large number of lattice spacings. A number of features 
emerge, all of which favor tadpole renormalization using mean links. 
This includes much better scaling of the hyperfine splittings in the 
three quarkonium systems. We also find that relativistic corrections to 
the spin splittings are smaller with mean-link tadpoles, particularly 
for the $c\bar c$ and $b\bar c$ systems. We also see signs of a breakdown in the NRQCD 
expansion when the bare quark mass falls below about one in lattice 
units (with the bare quark masses turning out to be much larger with 
mean-link tadpoles). 
\end{abstract}

\maketitle

\section*{}
Tadpole diagrams in lattice theories are induced by the nonlinear 
connection between the lattice link variables $U_\mu$ and the
continuum gauge fields. This causes large radiative 
corrections to many quantities in lattice theories. 
Most of the effects of tadpoles can be removed by a 
mean field renormalization of the links \cite{LepMac}
\begin{equation}
   U_\mu(x) \rightarrow {U_\mu(x) \over u_0} ,
\label{u0}
\end{equation}
where an operator dominated by short-distance fluctuations
is used to determine $u_0$. 

One of the earliest applications of tadpole improvement was in the 
development of lattice nonrelativistic quantum chromodynamics (NRQCD) 
\cite{LepThac,N1992,Nalphas,Nbmass,Nups,Npsi}. Precision simulations 
of the $\Upsilon$ system in NRQCD have provided important 
phenomenological results, including the strong coupling constant
\cite{Nalphas} and the $b$-quark pole mass \cite{Nbmass,Nups}. 
However the situation for charmonium is more problematic, due to
large relativistic corrections \cite{HDT}.

The quarkonium spectrum provides a powerful probe of 
tadpole renormalization. The quarkonium hyperfine 
spin splittings in particular are very sensitive to the details of 
the NRQCD Hamiltonian, with the relevant operators undergoing large 
tadpole renormalizations. For example, it has been shown \cite{HDT} 
that scaling of the charmonium hyperfine splitting is significantly 
improved when the tadpole renormalization is determined using 
the mean-link $u_{0,L}$ measured in Landau gauge 
\cite{LepMac,LepagePot,D234,NSglueball}:
\begin{equation}
   u_{0,L} \equiv
       \left\langle \mbox{ReTr} \, U_\mu \right\rangle, 
       \quad \partial_\mu A_\mu = 0 ,
\label{ulandau}
\end{equation}
compared to when the fourth root of the average plaquette $u_{0,P}$
is used:
\begin{equation}
   u_{0,P} \equiv
       \left\langle  \mbox{ReTr} \, U_{\mbox{pl}} 
       \right\rangle^{1/4} .
\label{uplaq}
\end{equation}

We make a comparison of the two tadpole
renormalization schemes $u_{0,L}$ and $u_{0,P}$ (implemented 
at tree-level) in the context of the quarkonium hyperfine splittings 
in NRQCD. This is done for the three quarkonium systems 
$c\bar c$, $b\bar c$, and $b\bar b$. The hyperfine splittings 
are computed both at leading ($O(M_Q v^4)$) and at next-to-leading 
($O(M_Q v^6)$) order in the relativistic expansion. 
(For further details see Ref.\ \cite{NSNRQCD})

All quantities are calculated after 
re-tuning of the lattice action parameters for each system.
 The resulting quark masses and lattice spacings for
the three quarkonium systems for the NRQCD action
at $O(v^6)$ are given in Tables \ref{TbaremL} and \ref{TbaremP}.
To minimize systematic errors from quenching 
the $b$-quark mass is tuned separately to ${}^1S_0$ for the 
$b\bar c$ and $b\bar b$ states with the $c$-quark mass tuned to the
${}^1S_0$  $c\bar c$ state.
Likewise, the lattice spacings were determined separately 
from the spin-averaged
$1P - 1S$ mass difference, which we set to 458~MeV (experimental 
value for charmonium).

\begin{table}[hbt]
\setlength{\tabcolsep}{0.5pc}
\caption{Bare quark masses for the three
quarkonium systems at $O(v^6)$, using Landau gauge mean-link 
tadpoles $u_{0,L}$; the stability parameter $n$ for each mass 
is given in square brackets.}
\label{TbaremL}
\begin{center}
\begin{tabular}{ccccc}
\hline
$\beta_L$  & $a_{c\bar c}$ (fm)
    & $aM_c^0[n]$   
    & $aM_b^0[n]$
    & $aM_b^0[n]$ \\
\hline
7.5  & 0.155(4) & 1.10[4]  & 3.20[2]  & 3.20[2] \\
7.4  & 0.179(2) & 1.20[4]  & 3.57[2]  & 3.57[2] \\
7.0  & 0.280(4) & 1.97[2]  & 6.10[2]  & 5.35[2] \\
6.85 & 0.319(5) & 2.25[2]  & 6.50[2]  & 5.90[2] \\
6.7  & 0.361(6) & 2.50[2]  & 7.20[2]  & 6.35[2] \\
6.6  & 0.380(7) & 2.67[2]  & 7.50[2]  & 6.66[2] \\
\hline
\end{tabular}
\end{center}
\vspace{-30pt}
\end{table}

\begin{table}[hbt]
\setlength{\tabcolsep}{0.5pc}
\caption{Bare quark masses for the 
quarkonium systems at $O(v^6)$ using average plaquette tadpoles 
$u_{0,P}$.}
\label{TbaremP}
\begin{center}
\begin{tabular}{ccccc}
\hline
$\beta_P$  & $a_{c\bar c}$ (fm)
    & $aM_c^0[n]$   
    & $aM_b^0[n]$
    & $aM_b^0[n]$ \\
\hline
7.3  & 0.140(4) & 0.65[8]  & 2.87[2]  & 2.87[2] \\
7.2  & 0.169(2) & 0.83[4]  & 3.20[2]  & 3.20[2] \\
7.0  & 0.210(2) & 1.10[4]  & 4.10[2]  & 3.95[2] \\
6.8  & 0.256(3) & 1.43[3]  & 4.98[2]  & 4.53[2] \\
6.6  & 0.313(4) & 1.80[3]  & 5.83[2]  & 5.23[2] \\
6.4  & 0.350(6) & 2.15[2]  & 6.45[2]  & 5.60[2] \\
6.25 & 0.390(6) & 2.41[2]  & 6.85[2]  & 5.99[2] \\
\hline
\end{tabular}
\end{center}
\vspace{-30pt}
\end{table}

The lattice NRQCD effective action for quarkonium is organized 
according to an expansion in the mean squared velocity 
$v^2$ of the heavy quarks, with corrections included 
for lattice artifacts. The effective action, including
spin-independent operators to $O(v^4)$, and spin-dependent 
interactions to $O(v^6)$, was derived in Ref.\ \cite{N1992}.
Following Refs.\ \cite{Nups,Npsi}, we use the evolution equation
\begin{equation}
   G_{t+1}\! = \!
    \left(\!1-\!\frac{aH_0}{2n}\!\right)^n 
   \!\!\!U^\dagger_4\!
   \left(\!1\!-\!\frac{aH_0}{2n}\!\right)^n\!\!\!
   \left(1\!-\!a\delta H\right) G_t .
\label{Gtp1}
\end{equation}

Relativistic corrections are organized in powers of the heavy 
quark velocity:
\begin{equation}
   \delta H = \delta H^{(4)} + \delta H^{(6)} .
\label{deltaH}
\end{equation}
Only next-to-leading spin-dependent interactions are considered in
$\delta H^{(6)}$  .

Simulations were done with 
the derivative operators and the clover fields corrected for their
leading discretization errors. 
Complete expressions for the operators can be found in
Refs. \cite{N1992,HDT,NSNRQCD}.

Gauge-invariant source and sink smearing was used for Meson  operators.

The gauge-field configurations were made using a tree-level 
$O(a^4)$-accurate tadpole-improved action \cite{LepCoarse}

\begin{figure}[htb]
\vspace{-30pt}
\caption{Hyperfine splittings with $u_{0,L}$ versus lattice 
spacing squared.}
\psfig{file=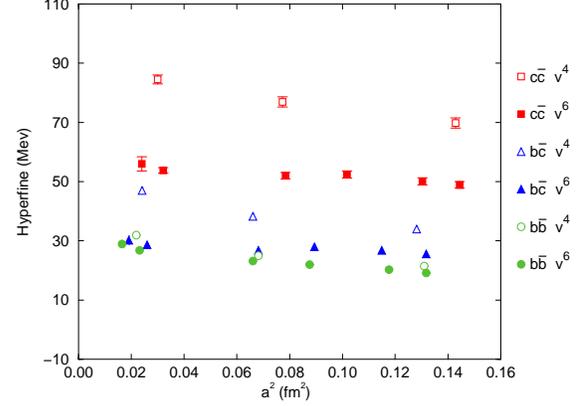,height=5.5cm,width=7.5cm}
\vspace{-18pt}
\vspace{-15pt}
\label{FHu0L}
\end{figure}

\begin{figure}[htb]
\vspace{-30pt}
\caption{Hyperfine splittings with $u_{0,P}$ versus lattice 
spacing squared.}
\psfig{file=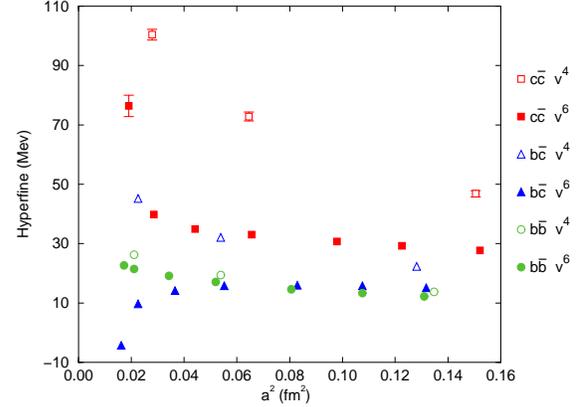,height=5.5cm,width=7.5cm}
\vspace{-18pt}
\label{FHu0P}
\end{figure}

There are a number of clear features in the data as shown in
Figures. \ref{FHu0L} and \ref{FHu0P}. 
To begin with, 
note that the results with $u_{0,L}$ 
 show smaller scaling violations than the results with 
$u_{0,P}$ .
 The smallest scaling violations are in 
$u_{0,L}$ at $O(v^6)$, which show little change over a 
large range of lattice spacing.
The scaling analysis provides evidence that $u_{0,L}$ tadpole 
renormalization yields a more continuum-like action than does $u_{0,P}$. 

The most striking feature is the drop in 
the $b\bar c$ splitting at smaller lattice spacings, when $u_{0,P}$ 
is used at $O(v^6)$ . 
Most  $c$-quark data with $u_{0,P}$ show  large 
changes at  small lattice spacings.
 The $u_{0,L}$ data exhibit much smoother behavior. 

We interpret these features as possible indicators of a breakdown 
in the NRQCD effective action at smaller lattice spacings, when the bare 
quark mass in lattice units $aM_Q^0$ falls below one. 
The bare $c$-quark mass is larger when $u_{0,L}$ 
is used.

Another key feature is that the relativistic
corrections to the hyperfine splittings are smaller 
when the action is renormalized using $u_{0,L}$. 
For example, we find that the charmonium hyperfine splitting is 
reduced by about 30--40\% in going from $O(v^4)$ to $O(v^6)$ 
when using $u_{0,L}$, compared to a reduction of about 40--60\% 
when using $u_{0,P}$. 
(Relativistic corrections have been analyzed 
in the $\Upsilon$ system \cite{Manke,Spitz}, and 
in heavy-light mesons \cite{ArifaHL,Ishikawa,Lewis}).
The $u_{0,P}$ 
relativistic corrections depend  strongly on the lattice 
spacing and may be related to the pathologies discussed above.
This casts new light on the results obtained in Ref.\ \cite{HDT},
where relativistic corrections to spin splittings in NRQCD were
first calculated. The velocity expansion for 
charmonium may not be as unreliable as was suggested.

We note finally that it is reasonable to attempt to extrapolate
the $O(v^6)$ hyperfine splittings for $c\bar c$ and $b\bar c$
to zero lattice spacing, from the data on coarse lattices, where 
there is good scaling behavior. However, the extrapolations 
in the $u_{0,L}$ and $u_{0,P}$ data are clearly very different.
 This suggests that 
some relevant operator coefficients $c_i$ in the NRQCD action 
receive significant $O(\alpha_s)$ 
corrections. This 
underlines the need to go beyond tree-level tadpole improvement 
in order to clarify the differences between
renormalization schemes. 

We have presented evidence that favors tadpole 
renormalization using the mean-link in Landau gauge over the fourth 
root of the average plaquette. This includes a demonstration of 
better scaling behavior of the hyperfine splittings in three 
quarkonium systems when $u_{0,L}$ is used, and a smaller size 
for spin-dependent relativistic corrections.
These results help to elucidate the structure
of the NRQCD effective action.

We are indebted to C.~T.~H. Davies, G.~P. Lepage, R. Lewis, 
and R.~M. Woloshyn for many helpful discussions and suggestions.
We also thank T. Manke for useful comments.
This work was supported in part by the 
Natural Sciences and Engineering Research Council of Canada.

\end{document}